# Texture transitions in binary mixtures of 6OBAC with compounds of its homologous series

A. Sparavigna[1], A. Mello[1], and B. Montrucchio[2]
[1] Dipartimento di Fisica, Politecnico di Torino
[2] Dipartimento di Automatica ed Informatica, Politecnico di Torino
C.so Duca degli Abruzzi 24, Torino, Italy

Recently we have observed in compounds of the 4,$n$-alkyloxybenzoic acid series, with the homologous index $n$ ranging from 6 to 9, a texture transition in the nematic range which subdivides the nematic phase in two sub-phases displaying different textures in polarised light analysis. To investigate a persistence of texture transitions in nematic phases, we prepared binary mixtures of 4,6-alkyloxybenzoic acid (6OBAC) with other members (7-,8-,9-,12-, 16OBAC) of its homologous series. Binary mixtures exhibit a broadening in the temperature ranges of both smectic and nematic phases. A nematic temperature range of 75°C is observed.
In the nematic phase, in spite of the microscopic disorder introduced by mixing two components, the polarised light optics analysis of the liquid crystal cells reveals a texture transition. In the case of the binary mixture of 6OBAC with 12OBAC and with 16OBAC, that is of compounds with monomers of rather different lengths, the texture transition temperature is not homogeneous in the cell, probably due to a local variation in the relative concentrations of compounds.

*Keywords*: Liquid crystals; Phase transitions; Alkyloxybenzoic acids

## 1. Introduction

Some liquid crystal compounds display mesophases because their molecules are able to form hydrogen-bonded dimers, rigid and long enough to provide mesogenic conditions. Among these materials, the series of alkyloxybenzoic acids and of cyclohexane-carboxylic acids, widely studied in the past, have recently gained a renewed interest because members of these families possess rather interesting features. In these liquid crystals in fact, the change induced in the dimer structure by an increase of temperature is giving phenomena such as the texture transitions in the nematic phase, the appearance of domains of spontaneous twist and the formation of dendrite structures [1-9].

According to the length of the alkyl tails, the alkyloxybenzoic acid possesses nematic or smectic, or both mesophases. In the nematic phase, the material is a mixture of monomers and dimers, with relative concentrations depending on the temperature [10]. In 4,$n$-alkyloxybenzoic acids ($n$OBAC), monomers are composed of two sterically distinct molecular parts, the oxybenzoic acid residue and the aliphatic chain (monomer unit $C_nH_{2n+1}OC_6H_{10}COOH$). Monomers form closed and open dimers, the molecular structures of which have been recently investigated in the framework of ab-initio calculations [11]. The number $n$ of carbon atoms in the aliphatic tail gives origin to the $n$OBAC series. Members with a number of carbon atoms in the tails ranging from 3 to 6 carbon atoms have a nematic but not a smectic phase; from 7 to 18 carbon atoms in the alkyl tails are smectogenic. The members with $7 \leq n \leq 13$ display both smectic C and N phases [12,13].

Recently optical investigations in compounds of the 4,*n*-alkyloxybenzoic acid series, with homologous index *n* ranging from 6 to 9, show a nematic phase subdivided in two sub-phases, characterised by different textures. Hereafter we denote with N' the nematic sub-phase below the texture transition temperature, and with N'' the nematic sub-phase above this temperature. Microscope observations show for the sub-phase N' a smectic-like texture [1-6]. The compound with 6 carbon atoms in the tail, 6OBAC, is rather interesting: it is not able to display a smectic phase but possesses the smectic-like nematic sub-phase N' in the nematic range, as 7-, 8- and 9OBAC [6].

The transition from N'' to N' is ascribed to the growth of cybotactic clusters [14-18] having short-range smectic order in the nematic phase under a certain temperature. This means that in the nematic melt, if the temperature is low enough, the closed dimers can aggregate in clusters with smectic C ordering. In N'', the local smectic order of the cybotactic clusters is destroyed by the high concentration of monomers and open dimers. Ultrasonic absorption measurements in nematic 7OBAC reveal the presence of relaxation processes, attributed to rearrangements of the equilibrium distributions of open dimers and closed dimers [19]: below the texture transition in the nematic phase, the ultrasonic absorption increases for the presence of bulk viscosity contributions. This is in agreement with the existence of cybotactic clusters in N'.

In order to investigate a persistence of the texture transitions in nematic phases, we used the 4,6-alkyloxybenzoic acid (6OBAC) as one of the components in binary mixtures with other members of the homologous series (7OBAC, 8OBAC, 9OBAC, 12OBAC and 16OBAC). We have also investigated the binary mixture of 6OBAC with the hexyl-cyclohexane carboxylic acid (hereafter called C6), a material with hydrogen bonded dimers too [5]. Table I reports the transition temperatures, determined with polarised microscope observations, of the compounds used for binary mixtures: the phase ranges of alkyloxybenzoic acids are also reported in Figure 1 for a more easy comparison. The alkyloxybenzoic compounds show crystal-crystal transitions in agreement with previous observations [12,20]. The crystal phase appearing on cooling from the smectic phase, in the compounds 6-, 7-, 8- and 9OBAC (but also in the binary mixtures of 6OBAC with 7-, 8- and 9OBAC here considered) has a texture composed of wide homogenous regions. It is possible to see in Fig.2 an example of what happens in cooling the 9OBAC acid.

Binary mixtures exhibit an increase in the temperature ranges of smectic and nematic phases, as it happens in other binary mixtures of alkyloxybenzoic acids [21,22]. In the binary mixtures, the mesogenic units are dimers of the same acid (homodimers) but also hydrogen bonded pairs of two different acids (heterodimers) [22]. In the nematic phase, in spite of the microscopic disorder introduced by mixing two components, the polarised light microscope analysis of the liquid crystal cells reveals the presence a texture transition. This means a persistence of cybotactic clusters, also in the case of mixed dimers. In the nematic of 6OBAC mixed with 7- or 8OBAC a well defined transition temperature exists. If 6OBAC is mixed with 9OBAC, this temperature slightly changes if different places of the liquid crystal cell are observed.

12OBAC does not possesses a texture transition in the nematic phase. If we take a glance at Figure 1, we see that the nematic phase of 12OBAC is narrow and at rather high temperature. The smectic-like nematic phase is then suppressed, included in a wide smectic phase. This is due to the fact that the compound has rather long dimers. 16OBAC does not possess a nematic phase. Binary mixture of 6OBAC with 12- and 16OBAC have huge nematic and smectic ranges: we observe a nematic temperature range of 75°C, mixing 6OBAC with 16OBAC. These mixtures show texture transitions in the nematic phase too, but with a texture transition temperature not homogeneous in the liquid crystal cell. In some places of the cell, it is not possible to identify the transition temperature, but we will discuss more deeply all these points in the following.

## 2. Binary mixtures of 6OBAC with 7-, 8-, 9-, 12-, 16OBAC and C6.

We have prepared the mixtures approximately 1:1 in weight, of 6OBAC with other members of the homologous series. The samples are 6OBAC-7OBAC (49:51), 6OBAC-8OBAC (52:48), 6OBAC-9OBAC (48:52), 6OBAC-12OBAC (53:47), 6OBAC-16OBAC (52:48). To the authors' knowledge these mixtures were not studied before. All the binary mixtures were inserted in the cell when the material was in the isotropic phase. The untreated glass surfaces of the cell walls were rubbed with cottonwool to favor a planar alignment. The liquid crystal cells were heated and cooled in a thermostage and textures observed with a polarized light microscope. To determine the temperature of texture transitions in the nematic phase, in some cases, we helped ourselves with the image-processing previously discussed in Ref. [6]. We have also prepared a binary mixture of 6OBAC with the trans-4-hexylcyclohexane carboxylic acid (monomer $C_6H_{13}C_6H_{12}COOH$), hereafter called C6. C6 displays a texture transition in the nematic phase too. In the binary mixture 6OBAC-C6 (47:53) we have not clear evidences of texture transitions. In the Table II, we report all the transition temperatures and in Figure 3 the temperature ranges of the crystal, smectic and nematic phases observed in the binary mixtures of alkyloxybenzoic acids for an easy comparison.

## 2.1. Binary mixtures with 7OBAC, 8OBAC and 9OBAC.

In Table II, we see the transitions temperatures of the binary mixture 6OBAC-7OBAC (ratio 49:51 in weight). A direct transition from the crystal phase in the nematic N' phase on heating is observed, but on cooling, a smectic phase appears between nematic and crystal. On heating then, the binary mixture behaves as pure 6OBAC, which does not posses a smectic phase, on cooling as 7OBAC which has a smectic phase. Figure 4 shows the nematic N'', N' sub-phases and the smectic textures on cooling: the nematic phase N'' texture turns into a smectic-like nematic phase N' texture.

In 6OBAC a rich variety of crystal phases was found [6]; we then prepared a mixture of 6OBAC and 7OBAC with the ratio 62:38 in weight. In this case, the mixtures has three crystalline textures on heating and two on cooling, with the following temperature transitions: CrI-CrII at 60°C, CrII-CrIII at 89°C, CrIII-N' starting at 95.5°C, N'-N'' at 125°C and N''-I at 148°C. The transitions on cooling are as follows: N''-N' at 114°C, N'-Sm at 89°C, Sm-Cr* at 83.5°C, and Cr*-CrI at 82°C. In fact, the crystal texture of Cr* is rather different from those of CrI, CrII and CrIII. As previously told, the crystal phase appearing on cooling from the smectic phase, in the binary mixtures of 6OBAC with 7-, 8- and 9OBAC, and in the pure compounds 6-, 7-, 8- and 9OBAC has a texture composed of wide homogenous regions (see Fig.2).

As reported in Table II, the binary mixture 6OBAC-8OBAC (52:48) displays an increase in the temperature range of the smectic and nematic range, on heating and on cooling. The texture transition is observed in the nematic phase at 129°C on heating and at 124°C on cooling.

The binary mixture 6OBAC-9OBAC (ratio 48:52 in weight) has a smectic range of ~30°, and a nematic range of ~55°, on heating and on cooling. It possesses a beautiful texture transition (Fig.5) in the nematic phase. Checking several spots of the liquid crystal cell, we noted that the mixture 6OBAC-9OBAC has a texture transition temperature different from place to place and comprised between 122°C and 128°C on heating, and between 124°C and 118°C on cooling. For the binary mixtures previously discussed (6OBAC-7OBAC and 6OBAC-8OBAC), the transition temperature is almost the same in the liquid crystal cell.

For this 6OBAC-9OBAC mixture, a texture transition temperature varying from place to place in the cell can be attributed to a local fluctuation of relative concentrations of homo- and heterodimers. Since the texture transition is due to the growth of cybotactic clusters in the nematic melt, a change in relative concentrations forces the cybotactic clusters to appear at different temperatures. Because monomers of 6- and 9OBAC have different lengths, a different viscosity can be supposed for the two materials, and then, when the mixture is inserted in the cell in the isotropic phase, local different concentrations of monomers are possible, turning then into a final non-homogeneous concentration of homo/heterodimers in the cell.

## 2.2. Binary mixtures with 12OBAC and 16OBAC.

The binary mixtures previously studied are composed of dimers with more or less the same length. Mixtures display wide nematic ranges and an increase of the smectic range with the increase of the monomer unit length in one of the mixture components. Now, let us see what happens when we mix components with rather different monomer lengths. Let us remember that 12OBAC and 16OBAC have so long monomers that smectic is favoured against the nematic phase.

As shown in Fig.3, the binary mixtures 6OBAC-12OBAC (53:47) has a nematic range of ~55°. On cooling from the isotropic phase, the cell remains in the smectic phase till 60°. Figure 6 displays the crystal, the smectic and the nematic phase of the mixture. For the mixture 6OBAC-16OBAC (52:48), the nematic range is so wide on cooling to be of ~75°. We observed just two transitions on heating from room temperature: a transition from a phase with a smectic-like texture into a nematic phase and from the nematic into the isotropic phase (see Fig.7). Then on cooling, we see the transition from the isotropic to the nematic phase and from nematic to smectic. The monomers of 6OBAC and 16OBAC are so different in length that a crystal phase in not seen at room temperature.

For both mixtures (6OBAC-12OBAC and 6OBAC-16OBAC), and quite surprisingly, a texture transition in the nematic phase can be observed. The texture transitions looks like those previously observed in the other binary mixtures (see Fig.8 and 9). But, near the side of the cell where the isotropic liquid of the mixture is inserted by capillarity, we were not able to identify a texture transition. In these regions, only one nematic texture (N''-like) is observed till the clearing point. We have then a temperature $T^*$ of the texture transition, function of the distance from the insertion side: increasing the distance, the smectic-like texture is observed and $T^*$ increases till the highest values of 130-134°C on heating.

As we have previously discussed for the mixture 6OBAC-9OBAC, local variations in the texture transition temperature can be due to non-homogeneous concentrations of homo- and heterodimers (closed and open) and monomers of the acids. A gradient in these concentrations could be induced by the flow of the isotropic material in the cell. If we consider 6OBAC monomers as more fluid for instance, a higher concentration of 6OBAC would be produced at the side of the cell opposite to the insertion one. In that place of the cell, the growth and persistence of cybotactic clusters is favored, because a high temperature $T^*$ is observed. Near the insertion side, the local smectic order is suppressed and the smectic-like nematic texture not observed, probably because of a high concentration of 12-,16OBAC monomers. Further experimental and theoretical investigations are necessary to understand if this explanation of the observed texture behavior in the nematic melt can be exhaustive or not.

**2.3. Binary mixture 6OBAC-C6.**
We have also prepared a binary mixture of 6OBAC with C6, a trans-4-hexylcyclohexane carboxylic acid (monomer $C_6H_{13}C_6H_{12}COOH$). This material possesses a smectic B phase and it is able to display a texture transition in the nematic phase too. In the Table II, we report the transition temperatures of the binary mixture 6OBAC-C6.
We have not seen a clear evidence of a texture transition in the nematic phase but a very smooth transition from a low temperature texture with several defects to a more homogenous texture. This could be explained with the two different local smectic orders of the two materials, smectic-B for C6 and smectic-C for 6OBAC. The mixture is not able to develop a local well defined order and then cybotactic clusters do not grow in the nematic melt.

**3. Conclusions.**
A complete investigation of the nematic and smectic ranges and of texture transitions for several weight ratios of the components in these binary mixtures was not the aim of this research and is devoted to future works. We were interested to investigate the possibility of texture transitions in the nematic phases of binary mixtures.
We observed in the binary mixtures with weight ratio approximately 1:1, that nematic and smectic ranges are strongly increased when materials with different monomer lengths are mixed. In the nematic phase, the melt is composed of monomers, and open and closed homo- and heterodimers: when the lengths of monomers of the two mixture components are approximately equal, a well defined texture transition in the nematic phase is observed. We can then conclude that cybotactic clusters with local smectic order appear under the texture transition temperature in the binary mixtures too. The texture of the nematic phase with cybotactic clusters has a smectic-like appearance: textures and transitions look like those observed in pure alkyloxybenzoic acids. Moreover, memory effects of schlieren lines in the nematic and smectic phase are observed during heating and cooling of cells filled with mixtures, as those observed in the cells filled with pure compounds.
When in mixtures the monomer lengths are rather different, the texture transition turns out to be suppresses near the side of the liquid crystal cell where the sample in its isotropic phase is inserted. The flow of the isotropic melt in the cell influences the effective concentrations of homo- and heterodimers and then the existence of cybotactic clusters.

TABLE I

| Compound | Transition temperatures (°C) |
|---|---|
| 6OBAC | Cr – 49 – Cr – 68 – Cr – 93 – Cr – 106 – N' – 129 – N'' – 147 – I<br>I – 147 – N'' – 127 – N' – 94 – Cr – 65 – Cr |
| 7OBAC | Cr – 92 – Cr – 94 – Sm – 99 – N' – 118 – N'' – 144 – I<br>I – 143 – N'' – 115 – N' – 96 – Sm – 89 – Cr – 78 – Cr |
| 8OBAC | Cr – 73 – Cr – 99 – Sm – 106 – N' – 130 – N'' – 143 – I<br>I – 142 – N'' – 124 – N' – 105 – Sm – 94 – Cr – 52 – Cr |
| 9OBAC | Cr – 93 – Cr – 95 – Sm – 117 – N' – 132 – N'' – 143 – I<br>I – 142 – N'' – 123 – N' – 114 – Sm – 90 – Cr – 64 – Cr |
| 12OBAC | Cr – 65 – Cr – 91 – Sm – 131 – N – 138 – I<br>I – 137 – N – 130 – Sm – 88 – Cr |
| 16OBAC | Cr – 93 – Sm – 133 – I<br>I – 132 – Sm – 88 – Cr |
| C6 | Cr – 32 – Sm – 47 – N' – 62 – N'' – 96 – I<br>I – 91 – N'' – 54 – N' – 42 – Sm |

**Table I: Transition temperatures (in °C) of the members of alkyloxybenzoic acid series used for the binary mixtures and of C6 (hexyl-cyclohexane carboxylic acid) on heating and on cooling. 12OBAC does not show a texture transition in the nematic phase. 16OBAC exhibits just a smectic phase. (Cr crystal; Sm smectic; N',N'' nematic subphases).**

TABLE II

| Compound | Transition temperatures (°C) |
|---|---|
| 6OBAC – 7OBAC | Cr – 96 – N' – 132 – N'' – 148 – I<br>I – 147 – N'' – 124 – N' – 87 – Sm – 83 – Cr – 80 – Cr |
| 6OBAC – 8OBAC | Cr – 72 – Sm – 86 – N' – 129 – N'' – 149 – I<br>I – 148 – N'' – 124 – N' – 83 – Sm – 67 – Cr – 53 – Cr |
| 6OBAC – 9OBAC | Cr – 60 – Sm – 89 – N' – (*) – N'' – 146 – I<br>I – 145 – N'' – (**) – N' – 86 – Sm – 55 – Cr |
| 6OBAC – 12OBAC | Cr – 72 – Sm – 96 – N – 142 – I<br>I – 140 – N – 92 – Sm – 60 – Cr |
| 6OBAC – 16OBAC | Sm – 88 – N – 143 – I<br>I – 142 – N – 65 – Sm |
| 6OBAC – C6 | Cr – 36 – N – 112 – I<br>I – 111 – N – 34 – Cr |

**Table II: Transition temperatures (in °C) in the binary mixtures of alkyloxybenzoic acids. In the case of mixture 6OBAC-9OBAC, the texture transition temperature in the nematic phase changes from place to place in the cell and it is comprised between 122°C and 128°C on heating (*) and 124°C and 118°C on cooling (**). For texture transitions in mixtures of 6OBAC with 12OBAC and 16OBAC, see text for the discussion.**

**FIGURE CAPTIONS**

Figure 1: Crystal, smectic and nematic ranges of alkyloxybenzoic acids 6-,7-,8-,9-,12- and 16OBAC used for mixtures.

Figure 2: The crystal phase Cr* appearing on cooling from the smectic phase has a texture composed of wide homogenous regions. In the figure, the smectic phase of 9OBAC (at 92°C, left) is followed by the crystal Cr* (at 88°C, middle), that a further decrease of temperature turns into another crystal phase (60°C, right). Crystal texture Cr* is observed in cooling 6-, 7- and 8OBAC and the binary mixtures of 6OBAC with 7-,8- and 9OBAC. The height-size of the image is 1 mm.

Figure 3: Crystal, smectic and nematic ranges of the binary mixtures of 6OBAC with alkyloxybenzoic acids (7-,8-,9-,12- and 16OBAC).

Figure 4: The binary mixture 6OBAC-7OBAC exhibits a texture transition in the nematic phase. The figure shows the transition on cooling: from the left the nematic N'' at high temperature (130°C), the smectic-like nematic phase at low temperature (103°C) and the smectic phase at 86°C. The height-size of the image is 1 mm.

Figure 5: The texture transition in the binary mixture 6OBAC-9OBAC on heating. On the left, the smectic-like texture N' that is replaced by the N'' texture, on the right. The image in the middle shows the transition.

Figure 6: The binary mixture 6OBAC-12OBAC exhibits crystal (left), smectic (middle) and nematic (right) phases.

Figure 7: The binary mixture 6OBAC-16OBAC exhibits smectic (left), smectic-like N' nematic (middle) and N'' nematic (right) phases.

Figure 8: The texture transition in the binary mixture 6OBAC-12OBAC: on the left of the image the smectic-like texture N' is growing in the sub-phase N''.

Figure 9: The texture transition in the binary mixture 6OBAC-16OBAC on cooling: on the left the schlieren texture in the nematic phase N'' and its smectic-like appearance in the nematic N'.

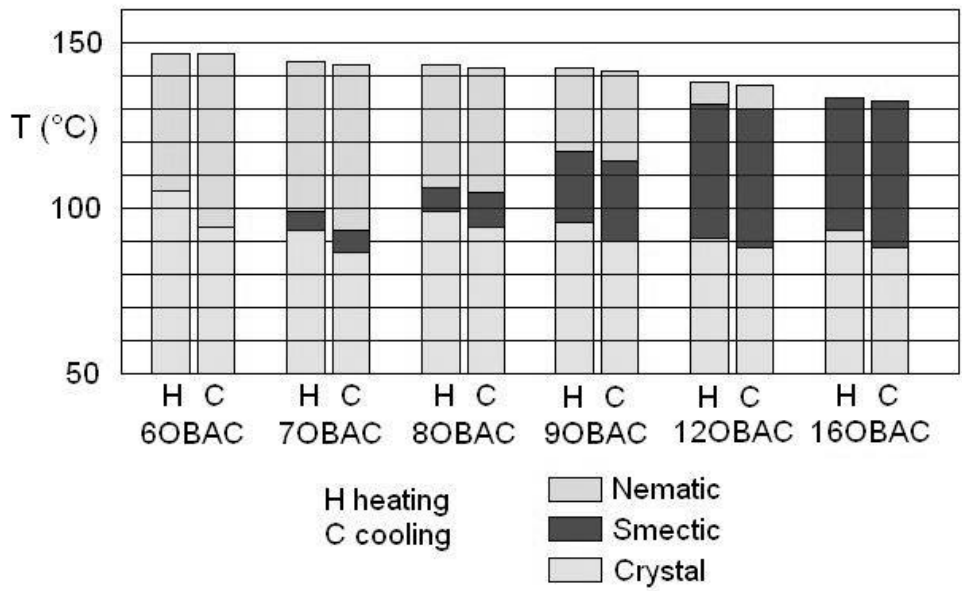

Figure 1

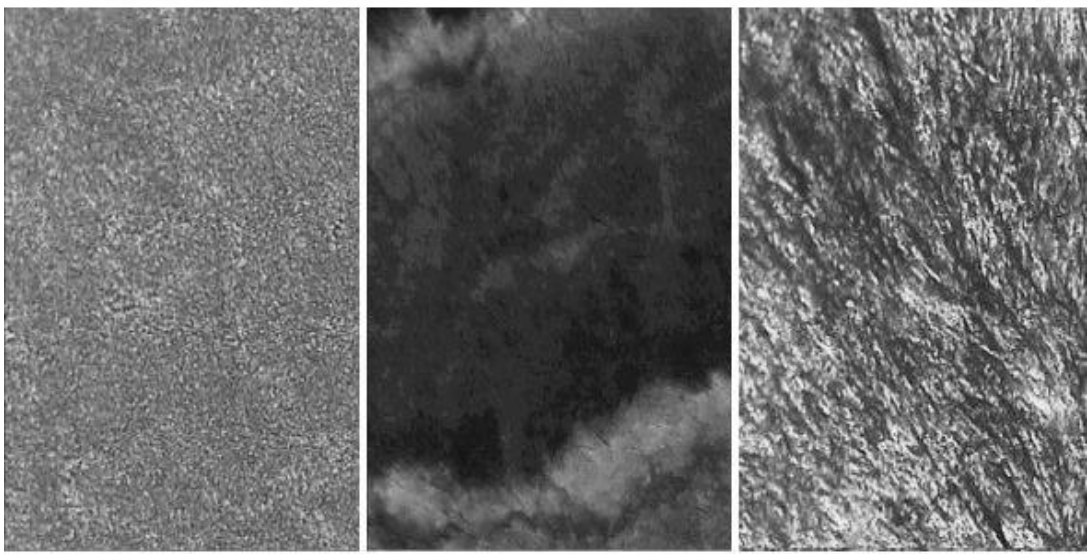

Figure 2

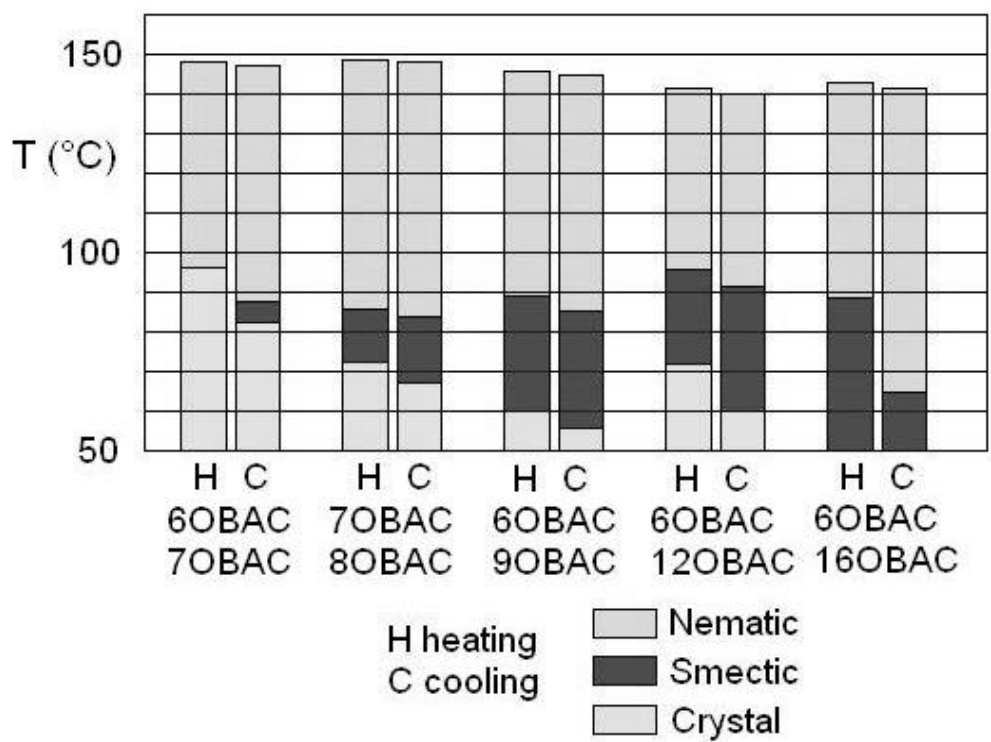

Figure 3

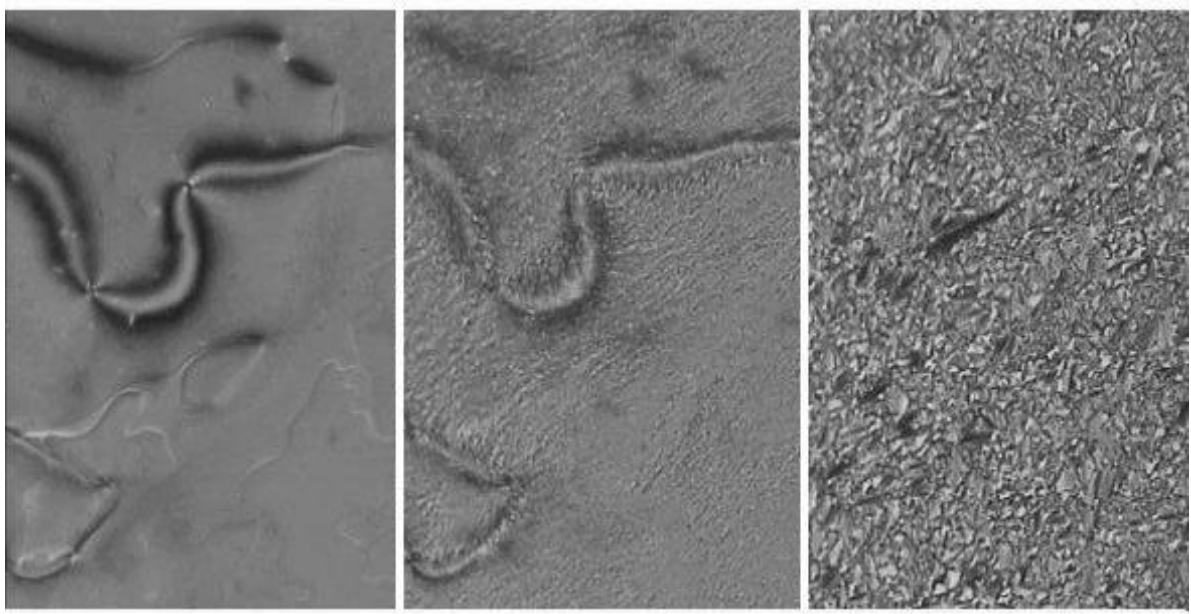

Figure 4

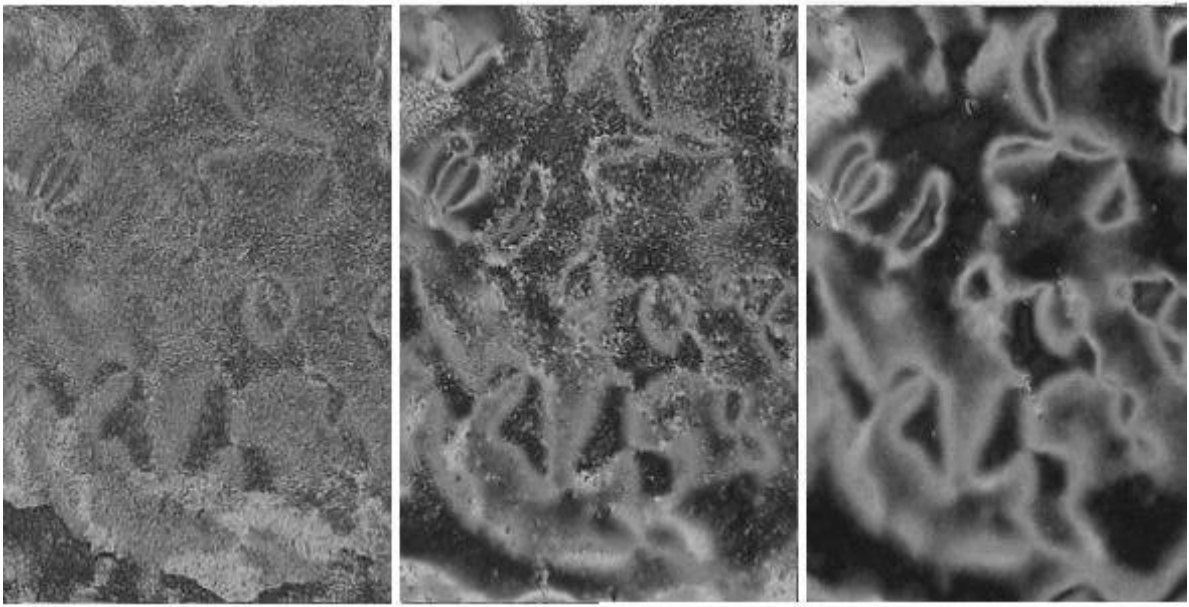

Figure 5

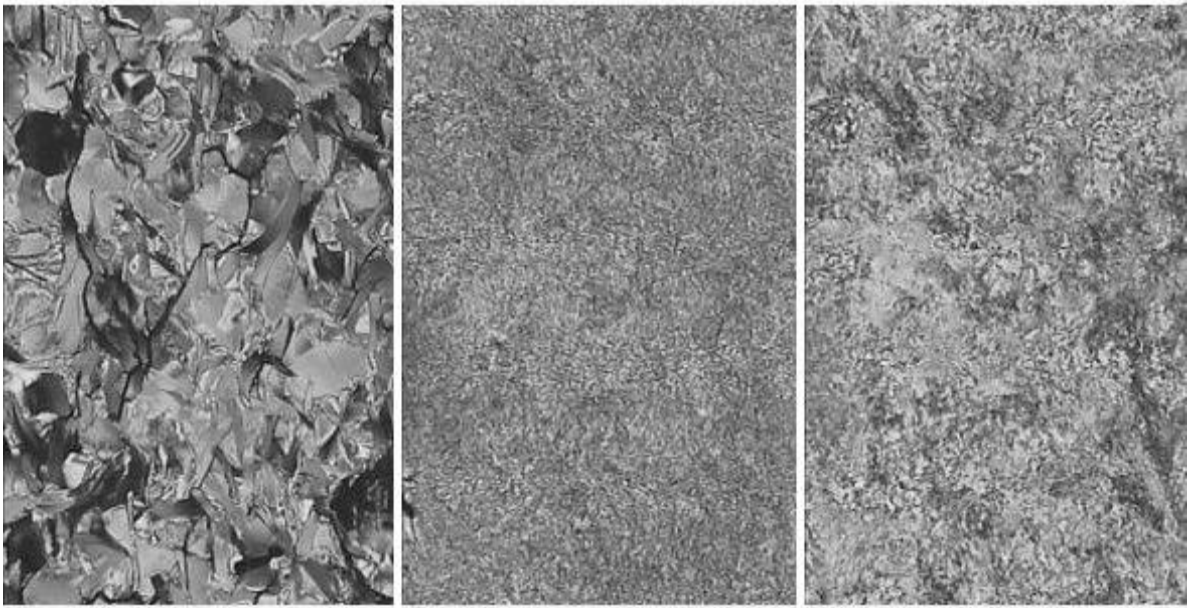

Figure 6

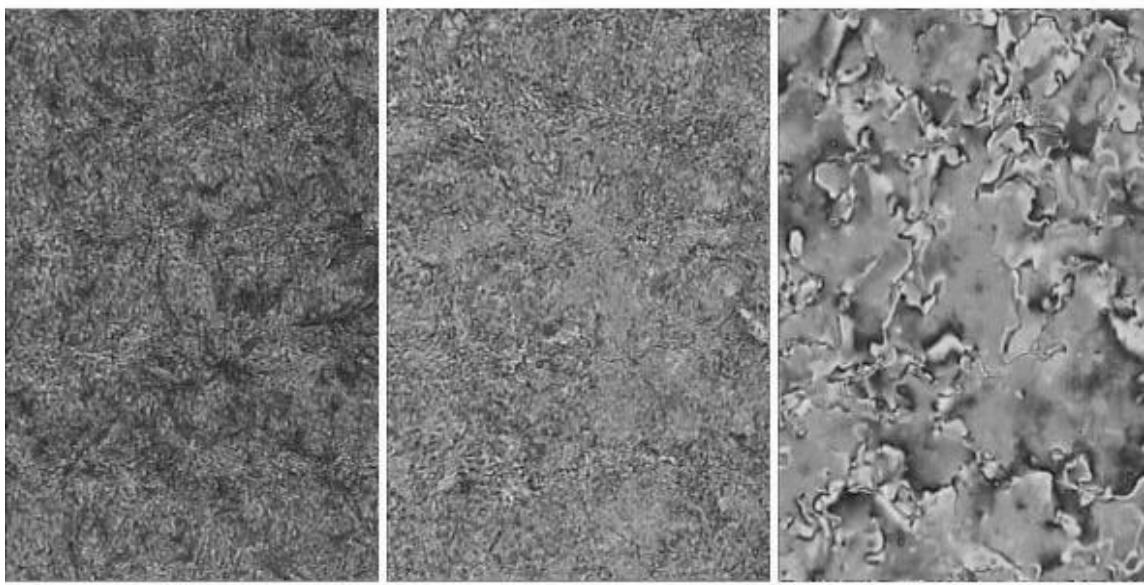

Figure 7

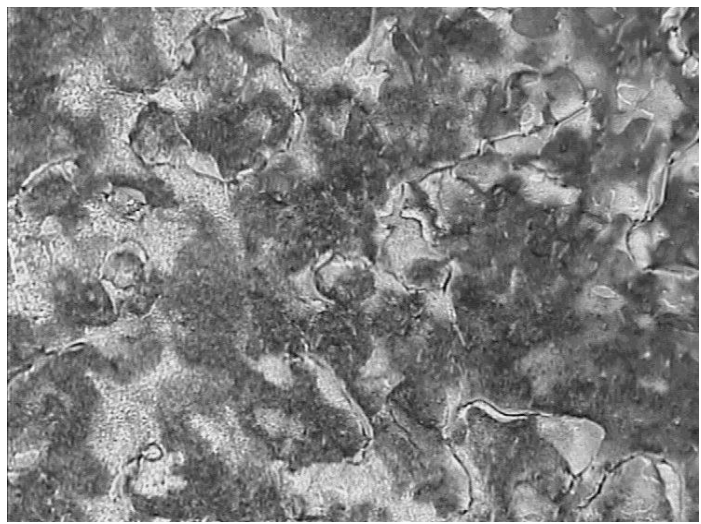

Figure 8

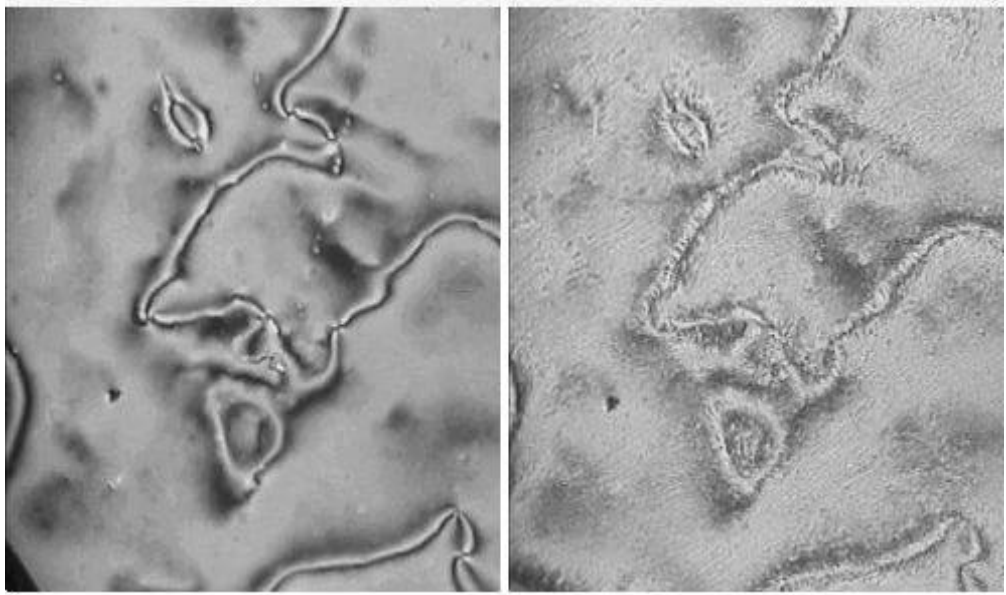

Figure 9